\documentclass{aa}
\usepackage{natbib,epsfig,txfonts,graphicx}
\bibpunct{(}{)}{;}{a}{}{,}

\def\Ha {H$\alpha$~}

\def\Hgama {H$\gamma$~}
\def\Hdelta{H$\delta$~}

\def \he {He {\sc i} $\lambda$5876\,}
 
\def\Mdot {$\dot M$\,}
\def\MV {M$_{\rm V}$\,}

\def\kms {km~s$^{\rm -1}$\,} 
\def\Rstar {$R_\star$\,}
\def\Mstar {$M_\star$\,}
\def\Teff {$T_{\rm eff}$\,}
\def\vinf {$v_\infty$\,}
\def\vesc {$v_{\rm esc}$\,}
\def\vsini {$v \sin i$\,}
\def\logg {$\log g$\,}

\def\Vr {$V_{\rm r}$\,}
\def\Vsys {$V_{\rm sys}$\,}

\def\logl {$\log L/L_\odot$\,}
\def\vmac {$v_{\rm mac}$\,}
\def\vmic {$v_{\rm mic}$\,}
 
\def \logwm {$log D_{\rm mom}$\,}

\def\Msun {$M_\odot$\,}
\def\Rsun {$R_\odot$\,}

\def \beq{\begin{equation}} 
\def  \eeq{\end{equation}}
\def \ben{\begin{enumerate}} 
\def \een{\end{enumerate}} 
\def \beqa{\begin{eqnarray}} 
\def \eeqa{\end{eqnarray}}
\begin{document}
\title{Wind structure of late B supergiants.}
\subtitle{I. Multi-line analyses of near-surface and wind
structure in HD~199\,478 (B8~Iae).}

   \author{N. Markova\inst{1}, R. K. Prinja\inst{2}, H. Markov\inst{1},
           I. Kolka\inst{3},
           N. Morrison\inst{4},
             J. Percy\inst{5},
           S. Adelman\inst{6}
}
   \offprints{N. Markova, \\ \email{nmarkova@astro.bas.bg}}

   \institute{Institute of Astronomy,   National Astronomical Observatory,
             Bulgarian Academy of Sciences, P.O. Box 136, 4700 Smoljan, Bulgaria \\
            \email{nmarkova@astro.bas.bg, hmarkov@astro.bas.bg}
        \and Department of Physics \& Astronomy, UCL, Gower Street, London
              WC1E 6BT, UK\\
	     \email{rkp@star.acl.uk}
	\and Tartu Observatory, Toravere 61602, Estonia \\
             \email{indrek@aai.ee}
	\and  Ritter Observatory, The University of Toledo, 
              Toledo, OH 43606, USA\\
	      \email{NMORRIS2@uoft02.utoledo.edu} 
        \and Department of Astronomy and Astrophysics, University of Toronto, Toronto ON Canada, M5S 3H4\\
	    \email{jpercy@utm.utoronto.ca}
	 \and  Department of Physics, The Citadel, Charleston SC 29409-0270, USA \\
	\email{adelmans@citadel.edu}
	}
\date{Received; accepted }

\abstract
{}
{
We provide a quantitative analysis of time-variable phenomena 
in the photospheric, near-star, and outflow regions of the late-B
supergiant (SG) HD~199\,478. This study aims to provide new 
perspectives on the nature of outflows in late-B SGs and on the 
influence of large-scale structures rooted at the stellar surface.}
{The analysis is  based primarily on optical spectroscopic 
datasets secured between 1999 and 2000 from the Bulgarian NAO, 
Tartu, and Ritter Observatories. The acquired time-series  samples 
a wide range of weak metal lines, He~{\sc i} absorption, and both 
emission and absorption signatures in H$\alpha$. Non-LTE line synthesis 
modelling is conducted using FASTWIND for a strategic set of 
late-B SGs to constrain and compare their fundamental parameters 
within the context of extreme behaviour in the \Ha lines.}
{The temporal behaviour of HD~199\,478 is characterised by three 
key empirical properties: (i) systematic central velocity shifts 
in the photospheric absorption lines, including C~{\sc ii} and 
He~{\sc i}, over a characteristic time-scale of $\sim$ 20 days;
(ii) extremely strong, variable \Ha emission with no 
clear modulation signal, and (iii) the occurrence in 2000 of a 
(rare) high-velocity absorption (HVA) event in H$\alpha$, which 
evolved over $\sim$ 60 days, showing the clear signature of  
mass infall and outflows. In these properties HD~199\,478  
resembles few other late-B SGs with peculiar emission and HVAs 
in H$\alpha$ (HD~91\,619, HD~34\,085, HD~96919). Different 
possibilities accounting for the phenomenon observed are 
indicated and briefly discussed.}
{At the cooler temperature edge of B SGs, there are objects 
whose wind properties, as traced by H$\alpha$, are inconsistent 
with the predictions of the smooth, spherically symmetric wind 
approximation. This discordance is still not  
fully understood and may highlight the role of a non-spherical, 
disk-like, geometry, which may result from magnetically-driven 
equatorial compression of the gas. Ordered dipole magnetic 
fields may also lead to confined plasma held above the stellar 
surface, which ultimately gives rise to transient HVA events.}


\titlerunning{}
\authorrunning{N. Markova et al.}

\maketitle

\section{Introduction}

The key limiting assumptions incorporated within current hot 
star model atmospheres include a globally stationary and 
spherically symmetric stellar wind with a smooth density 
stratification. Although these models are generally quite 
successful in describing the overall wind properties, there
are numerous observational and theoretical studies which 
indicate that hot star winds are certainly not smooth and 
stationary. Most (if not all) of the time-dependent constraints 
refer, however, to O-stars and early B supergiants (SGs), while 
mid- and late-B candidates are currently under-represented in 
the sample of stars investigated to date. 

 \citet{Vink00} have shown that for mid and late-B SGs there 
is a discrepancy between theoretical predictions and 
\Ha mass-loss rates, derived by means of unblanketed
model analysis. This finding was confirmed by recent 
investigations using line-blanketed model atmospheres 
(see, e.g., \citealt{crowther06, MP}). The reason for the 
discrepancies is not clear yet but wind structure and 
variability might in principle cause them.  Indeed, 
observations indicate that while winds in late-B SGs 
are significantly weaker than those in O SGs 
(e.g. \citealt{MP}), there is no currently established 
reason to believe that weaker winds might be less 
structured/variable than stronger ones (e.g. 
\citealt{Markova05, puls06}).

The first extended spectroscopic monitoring campaigns of 
line-profile variability ($lpv$) in late-B SGs were performed 
by \citet{kaufer96a, kaufer96b}, who showed that stellar winds 
at the cooler temperature edge of the B-stars domain can also be 
highly variable. Interestingly, in all 3 cases studied by these 
authors, the variability patterns (as traced by H$\alpha$) were 
quite similar consisting of (i) blue- and red-shifted emission with 
$V/R$ variations  similar to those in Be-stars, and (ii) the sudden 
appearance of deep and highly blue-shifted absorptions (HVAs). 
Though the kinematic properties of the HVAs in \Ha were found 
to be completely different from those of DACs (Discrete Absorption 
Components) in the UV spectra of O and early-B stars (e.g., 
HVAs do not propagate outwards, but instead extend to zero velocity 
and even indicate mass infall), similar scenarios consisting of 
large-scale wind structures rooted in the photosphere were suggested  
to interpret their appearance and development in time.

The present paper is focused on a multi-line investigation of
HD~199\,478 (HR~8020), a B8 Iae star whose stellar and wind 
properties have been recently determined by means of NLTE model 
atmosphere analysis of strategic lines in the optical \citep{MP}. 
The first extensive monitoring campaign of $lpv$ in the optical 
spectrum of this star revealed \Ha variability similar to those 
described by Kaufer et al. (i.e., peculiar wind emission with 
$V/R$ variations similar to those in Be stars) with one exception,
i.e. no indications for any HVAs were found during that survey 
\citep{MV}. 
\begin{table*}[t]
\caption{Spectral observations and instruments employed.}
\label{log1}
\begin{tabular}{llllllll}
\hline 
\hline
\multicolumn{1}{l}{Observatory}
&\multicolumn{1}{l}{}
&\multicolumn{1}{l}{Telescope}
&\multicolumn{1}{l}{Instrument}
&\multicolumn{1}{l}{{$R$=$\lambda/\delta\lambda$}}
&\multicolumn{1}{l}{window [in \AA]}
&\multicolumn{1}{l}{region}
&\multicolumn{1}{l}{N$_{\rm spec}$}
\\
\hline
\\
National Astron. Obs.  &NAO  &2.0m  &coud\"e    &15\,000  &204& \Ha &45\\
National Astron. Obs.  &NAO  &2.0m  &coud\"e    &30\,000  &100& \Ha &4\\
National Astron. Obs.  &NAO  &2.0m  &coud\"e    &15\,000  &204& \he &46\\
Tartu Observatory      &TO   &1.5m  &long-slit  &16\,000  &127& \Ha & 9\\
Ritter Observatory     &RO   &1.1m  &fiber-fed  &26\,000  & 70& \Ha & 7\\
\hline
\end{tabular}
\end{table*}
In addition to wind variability significant absorption $lpv$
was also established,
which raised suggestions of a link stellar pulsations. 
To check the pulsational hypothesis however long-term photometric 
observations were required.

Motivated by the intriguing time-variable properties reported
above, we organised and conducted new parallel spectroscopic and 
photometric monitoring campaigns of HD~199\,478 during 1999 
and 2000.  Details of the spectroscopic analysis are presented 
here and the results of the photometric survey were recently published 
by \citep{PAM}.

\section{Observations and data reduction}
\label{obs}

\subsection{Spectral data}

Spectroscopic data consisting of 65 spectra centered on \Ha and 
46 on \he  were predominantly collected at the National 
Astronomical Observatory (NAO), Bulgaria while individual \Ha 
observations were also secured at the Tartu Observatory (TO), 
Estonia and at Ritter Observatory (RO), USA. 

The total time coverage of these data is from January 1999 to December 
2000 with large data gaps in the summer and the winter each year. The 
time sampling was typically 3 to 6 spectra per month with a time-interval 
between successful exposures of 1 to 2 days, except for the fall of 
2000 when HD~199\,478 was monitored more intensively. The time distribution 
of the data and spectral regions observed are given in Table~\ref{log2} 
while specific information about the equipment, reduction strategy 
and methods used at each observatory is outlined below and summarised 
in Table~\ref{log1}.

\paragraph{National Astronomical Observatory, Smolyan, Bulgaria} $-$ A 
total of 95 high-quality spectra (49 in \Ha and 46 in \he) were 
obtained (observers T. Valchev and H. Markov) in the coud\'e focus 
of the NAO 2m telescope of the Bulgarian Academy of Sciences during 
the period March, 1999 -- December, 2000. Most of the observations 
(91) were carried out using a BL632/14.7 grooves $mm^{-1}$ grating 
in first order together with a PHOTOMETRICS CCD (1024 x 1024, 24$\mu$) 
as a detector \footnote{This detector was characterised by a $rms$ 
read-out noise 3.3 electrons per pixel (2.7 ADU with 1.21 electrons 
per ADU)}. This configuration produces spectra with a reciprocal 
dispersion of $\sim$0.2 \AA\ pixel$^{-1}$ and an effective resolution 
of $\sim$ 2.0 pixels ($\sim$0.44 \AA\, R=15\,000) over a wavelength 
range of $\sim$204 \AA\,. In addition, four more \Ha spectra with a 
reciprocal dispersion of $\sim$0.1 \AA\, pixel$^{-1}$ and an 
effective resolution of $\sim$ 2.0 pixels ($\sim$0.2 \AA\, R=30\,000) were 
obtained using a BL632/22.3 grooves $mm^{-1}$ grating in second 
order and the same detector. The S/N ratios of the NAO spectra range between 
250 and 450.

We followed a standard procedure for data reduction (developed in 
IDL) including: bias subtraction, flat-fielding, cosmic ray 
removal, wavelength calibration, correction for heliocentric radial 
velocity (\Vr = -12~\kms), water vapour line removal and rebinning to 
a step of 0.2 \AA\, per pixel. More information about the reduction 
procedure can be found elsewhere \citep{MV, Markova04}.

\paragraph{Tartu Observatory, Toravere, Estonia} $-$
At TO (observer I. Kolka) 
9 \Ha spectra were obtained using a 1.5-m reflector equipped 
with a Cassegrain spectrograph and an Orbis-1 (LN2-cooled) CCD camera 
(512x512 pixels). About 130~\AA\, is covered by one exposure with a 
reciprocal dispersion of ~0.25~\AA/pix resulting in a resolution of 
about 16\,000 at \Ha. 

The observations were reduced in a uniform way using MIDAS. Due to 
the very low dark current, the mean background subtracted from the 
raw frames is a sum of bias plus dark and real sky. Flat-fielding 
was not performed due to the reasonably flat response of the CCD 
and to the empirical result that the spectrum summed over 4 - 6 CCD 
rows has almost no distortion from pixel to pixel sensitivity 
differences. Thus, the photon noise and the read-out noise are the 
main sources of errors. The telluric water vapour lines were removed 
by dividing individual spectra with a scaled model telluric spectrum. 
Finally, the spectra were corrected to the stellar rest frame for a 
radial velocity of -12~\kms and normalised to the continuum.

\paragraph{Ritter Observatory, Toledo, USA} $-$ The Ritter spectra were 
obtained with a 1-m telescope, fiber-fed \'echelle spectrograph, 
and Wright Instruments CCD camera at the University of Toledo over 
a period of two months (September -- November) in 2000. The spectral 
resolving power $R\simeq26\,000$, with the resolution element having 
a FWHM of about 4 pixels. The spectral coverage consists of 9 separate 
70-\AA\ regions in the yellow and red. All the exposures were 1 hour 
in duration.
\begin{table}[t]
\caption{Summary of the spectral data sets.}
\label{log2}
\tabcolsep1.0mm
\begin{tabular}{llcrl}
\hline
\hline
\multicolumn{1}{l}{Region}
&\multicolumn{1}{c}{Observational dates}
&\multicolumn{1}{c}{HJD 2451200+}
&\multicolumn{1}{c}{N$_{\rm spec}$}
&\multicolumn{1}{c}{S/N}
\\
\hline
\\
\Ha + CII  & 1998 Dec.,30  -- May, 2      & 40.6 -- 101.6   & 8  & 240\\
\Ha + CII  & 1999 Sept.,17 -- Dec., 2     &239.4 -- 315.2   &15  & 423\\
\Ha + CII  & 2000 March, 28 -- June, 23   &432.6 -- 519.4   &10  & 275\\
\Ha + CII  & 2000 Sept, 5   -- Dec., 8    &592.7 -- 687.2   &32  & 320\\
\he        & 1999 March, 2  -- April, 24  & 40.6 --  93.5   & 4  & 240\\
\he        & 1999 Sept, 17  -- Dec., 2    &239.4 -- 315.2   &15  & 423\\
\he        & 2000 March, 28 -- June, 23   &432.6 -- 519.4   &10  & 275\\
\he        & 2000 Sept., 14 -- Dec., 8    &602.3 -- 687.2   &17  & 320\\
\hline
\end{tabular}
\end{table}

The raw frames were reduced with Ritter Observatory's standard 
reduction script under Sun/IRAF 2.11.3. \footnote{IRAF is 
distributed by the National Optical Astronomy Observatories, 
which are operated by the Association of Universities for Research 
in Astronomy, Inc., under contract with the National Science 
Foundation.} Removal of the telluric lines was done with the 
IRAF task {\em telluric}. The template spectra used in this context were 
artificial rows of Gaussians constructed from spectra of telluric 
standard stars taken under various conditions. For each spectrum 
of HD~199\,478, the template that provided the best telluric 
correction was used. However, for a few spectra none 
of the templates in the library were completely successful at removing 
telluric lines. The spectra were then Doppler corrected to the 
heliocentric rest frame and normalised to the continuum.

\subsection{Consistency check}

An important point of any study which relies on observations collected 
at various observatories, with different instruments and equipment, is 
the mutual consistency among the corresponding datasets. The ideal 
way to perform a consistency check is to compare strictly simultaneous 
data collected from different places. 

Fortunately, our sample has three such spectra, that were taken
at the NAO, TO and RO within 8 hours in the same night 
(Sept. 17, 2000). Using these spectra we checked for possible 
systematic differences in continuum and wavelength calibrations. 
The results obtained indicate that the wavelength calibration of 
the NAO and the TO spectra agree perfectly while the Ritter 
spectrum shows a one pixel systematic shift to the red. 

On the other hand, and as regards photometric calibration,
the Ritter spectrum (corrected for the shift of one pixel) 
fits quite well (within the noise) the NAO spectrum, while the 
relative fluxes between 6561 to 6584 \AA\, in the TO spectrum 
are up to 3\% stronger. Since the three spectra are  not 
strictly simultaneous and since line profile variations on a 
shorter (hours) time scale cannot be excluded, it is not
currently possible to judge to what extent 
the established differences in the fluxes redward of 
the emission peak of \Ha might be caused by imperfect continuum 
rectification or are due to real variability in the wind. 

Thus, differences of $\sim$9 \kms in velocity scale and up 
to 3\% in relative flux cannot be excluded in our spectral 
time-series (but see next section).

\section{Photospheric variability}
\subsection{Photometric evidence}
\label{phot_var}

Recent results \citep{PAM} indicate that the photometric behaviour 
of HD~199\,478 is characterised by continuous 
irregular/multi-periodic variations with an amplitude of about 
0.15~mag on a time-scale of 20 to 50 days. In some observational 
runs colour variations of up to 0.05~mag, in phase with the light
curve, have been also observed while in others no colour variations were 
detected above the corresponding error. In these properties 
HD~199\,478 is similar to other OB SGs which are known to be 
photometrically variable and show small amplitude microvariations 
in the visual, with little colour variations, on a time scale 
from days to months (see e.g. \citealt{Aerts99,VanG,M}).

\subsection{Photospheric variability traced by 
C II doublet and He I 6678 lines.}
\label{he1_c2}

\noindent
Following \citet{MV}, we used the absorption lines of C~{\sc ii}~$\lambda\lambda$6578.03, 6582.85, He~{\sc i}~$\lambda$6678.15 
and He~{\sc i}~$\lambda$5875.67 
to probe the deep-seated variability and photospheric structure 
of HD~199478 during the period covered by our observations. 
To improve the internal consistency of the 
wavelength scale in the extracted spectra from different 
observatories, which is of crucial importance for the 
purposes of the time-series analysis, the C~{\sc ii} and 
He~{\sc i} $\lambda$6678 line profiles were realigned using 
the diffuse interstellar band at $\lambda$6613.6 as a 
fiducial. Similarly the interstellar line of 
Na~{\sc i} D~$\lambda$5889.95 was aligned for our study of 
profile changes in He~{\sc i} $\lambda$5876. Following 
these adjustments,  we estimate that the velocity scale 
local to each line profile is stable to 1$-$2 km s$^{-1}$. 
We are also confident that the C~{\sc ii} lines are not 
severely affected by large fluctuations in the outer red 
wing of H$\alpha$.  For the photospheric analyses the 
C~{\sc ii} lines were normalised to a local continuum 
assigned (using a low-order polynomial) between 
$\lambda\lambda$6570 to 6590{\AA}.\\

Radial velocities of the selected He~{\sc i} and C~{\sc ii} 
lines were measured by fitting Gaussian profiles in the 
knowledge that these lines are generally very symmetric. 
For the 2000 datasets the following estimates 
of the mean radial velocity and peak-to-peak amplitude were 
derived: $-$1.9$\pm$3.9~km s$^{-1}$ and 16~km s$^{-1}$ for 
C~{\sc ii}~$\lambda$6578; $-$2.0 $\pm$3.7~km s$^{-1}$ and 
12~km s$^{-1}$ for He~{\sc i}~ $\lambda$6678 and  
+2.8$\pm$3.7 km s$^{-1}$ and  13 km s$^{-1}$ for 
He~{\sc i}~$\lambda$5876. Variations of $\sim$~15{\%} in the total 
equivalent widths of the lines 
were also established. There is a tighter correlation between 
the strength and velocity changes seen in C~{\sc ii} and 
He~{\sc i} $\lambda$6678 than between either of these lines 
and He~{\sc i} $\lambda$5876.

\begin{figure}
\begin{minipage}{8.8cm}
\resizebox{\hsize}{!}
{\includegraphics{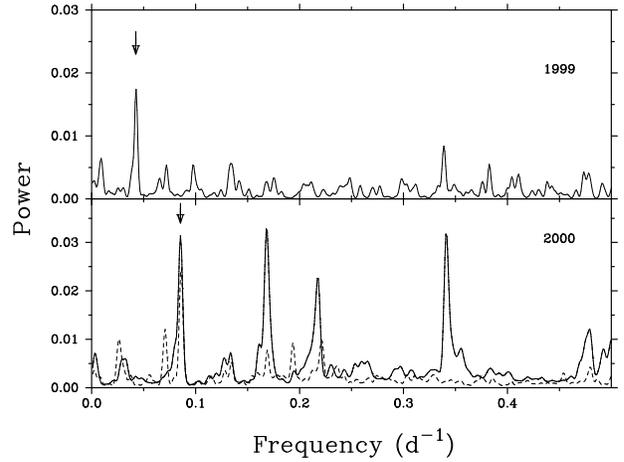}}
\end{minipage}
\hfill
\caption{Power spectrum (in arbitrary units) for the
C~{\sc ii} (solid line) and He~{\sc i} $\lambda$6678 photospheric
absorption lines. The arrow marks the `stable' peaks at
$\sim$ 11.7 days and 23.4 days.
}
\label{dft}
\end{figure}

The sampling rate of the 2000 (and the 1999) dataset is rather
uneven and short-time series secured over a few days are separated
by data gaps of between 1 to 3 months. This makes the search
for periodic signals rather more uncertain. We applied the CLEAN 
method \citep{Roberts87} to the radial
velocity measurements (using a gain of 0.5 and 200 iterations).
The power spectra, where the features of the window function have
are deconvolved using the discrete Fourier Transform, are shown
in Fig~\ref{dft}, for the 1999 C{\sc ii} and the
2000 He~{\sc i}	$\lambda$6678 and C~{\sc ii} data.

\begin{figure}
\begin{minipage}{7.8cm}
\resizebox{\hsize}{!}
{\includegraphics{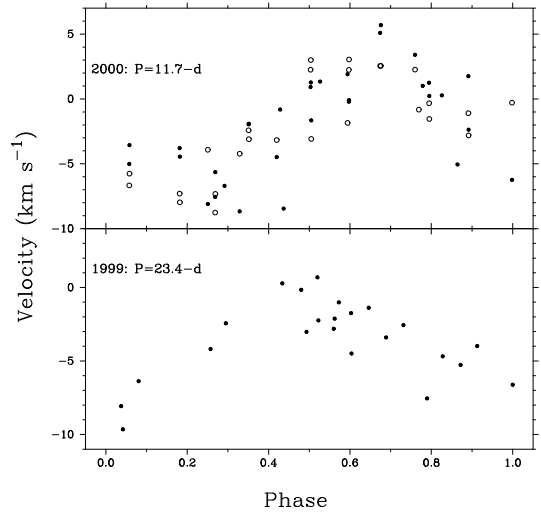}}
\end{minipage}
\hfill
\caption{The absorption central velocities of C~{\sc ii} (filled 
circles) and He~{\sc i} $\lambda$6678 (open circles) phased on 
the modulation time-scales identified in the 1999 and 2000 
datasets.
   }
   \label{vrad}
   \end{figure}

Clearly, there is no strictly periodic signal present in 
the photospheric lines of HD~199\,478 that remains coherent 
between 1999 to 2000. There is instead some indication that the 
absorption lines are semi-modulated in their central velocities 
over time-scales of $\sim$ weeks to months. The only signal in 
the 2000 power spectrum that is consistent between C~{\sc ii} 
and He~{\sc i} $\lambda$6678 is at a frequency of $\sim$~0.085 
days$^{-1}$, i.e. a period of $\sim$ 11.7 days. Interestingly, 
in 1999 this modulation is essentially absent, but the strongest 
peak in the C~{\sc ii} dataset at 0.0478 days$^{-1}$ corresponds 
to precisely twice the 11.7 days period.  We find no evidence for 
a 11.7 days or 23.4 days modulation in He~{\sc i}$\lambda$5876.
However, note that the 1999 data sampling is more fragmented, 
with a gap of around 140 days. 
The central velocities of the C~{\sc ii} and He~{\sc i}
$\lambda$6678 absorption lines phased on both these periods 
are shown in Fig~\ref{vrad}.  

Despite the high signal-to-noise and
spectral resolution of our data (Sect. 2.2) there is also no
evidence for sub-features travelling blue-to-red (prograde) 
in the absorption troughs of the lines, that might for example
be identified in terms of the presence of low-order non-radial
pulsations.

\subsection{Comparison between spectral and photometric
variability}

The 2000 differential $uvby$ photometry obtained by 
SA as well as the 2000 $UBV$ data collected by 
JP, though not strictly simultaneous, cover the same 
time period as the corresponding spectroscopic data. 
The Fourier and self-correlation analysis of these data 
indicate the presence of a periodic variation of 
18$\pm$4 ($UBV$) to 21$\pm$4 ($uvby$) days 
with an amplitude of about 0.15~mag.  The colour curve of this 
micro-variation is blue at the maxima and red at the 
minima of the light curve, thus resembling $\alpha$ Cyg 
variations in BA SGs.
\begin{figure*}
\begin{minipage}{5.8cm}
\resizebox{\hsize}{!}
{\includegraphics{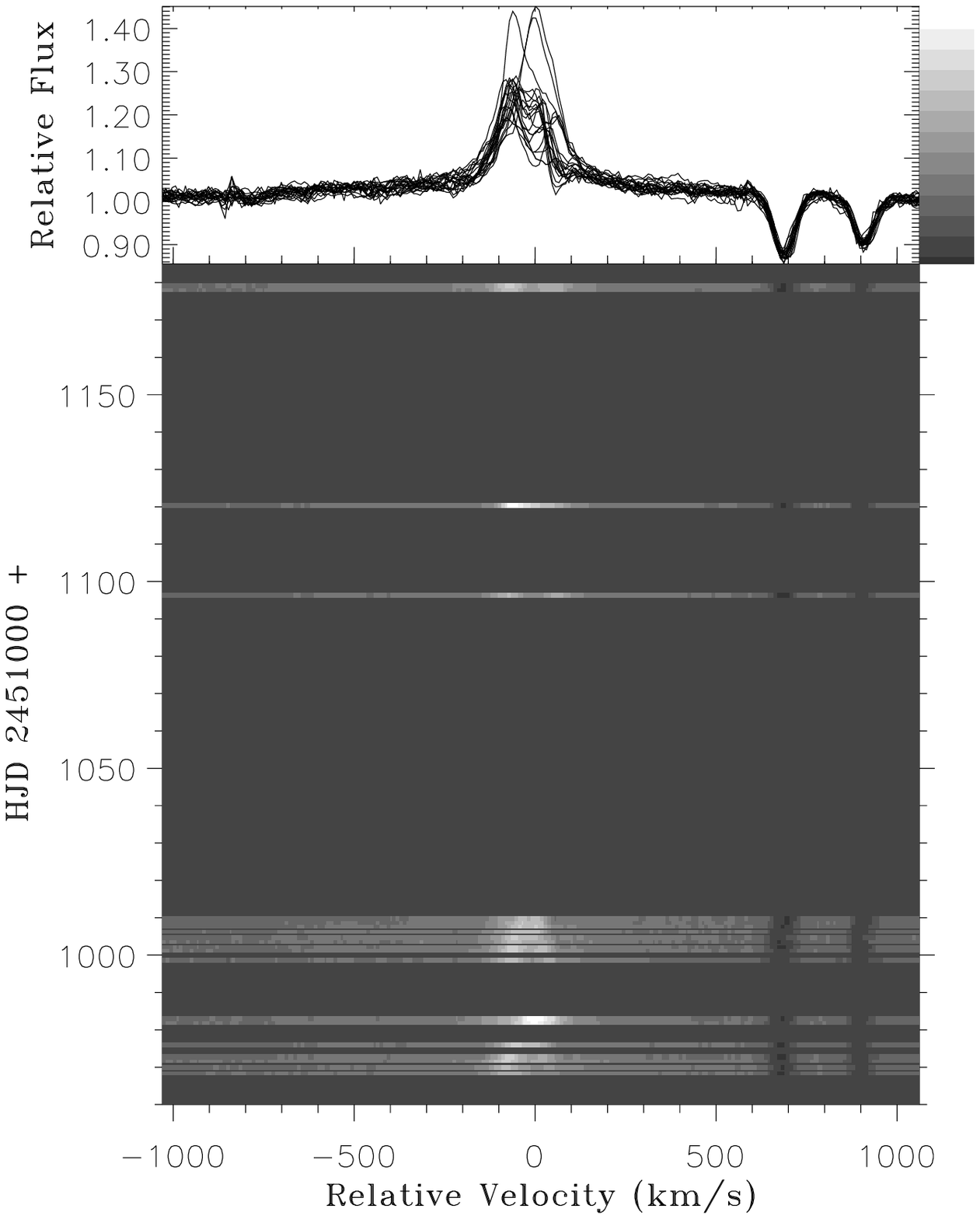}}
\end{minipage}
\hfill
\begin{minipage}{5.8cm}
\resizebox{\hsize}{!}
{\includegraphics{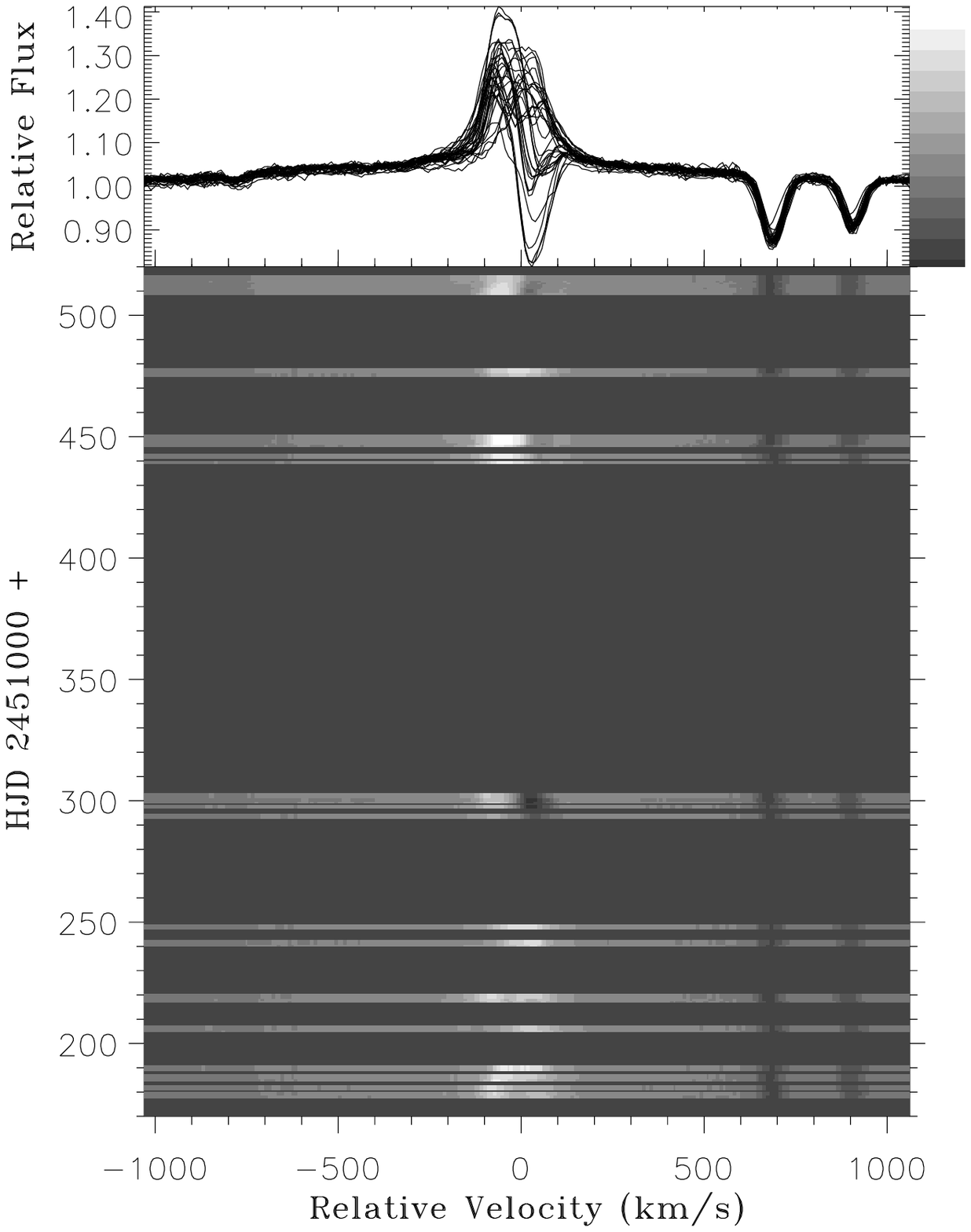}}
\end{minipage}
\hfill
\begin{minipage}{5.8cm}
\resizebox{\hsize}{!}
{\includegraphics{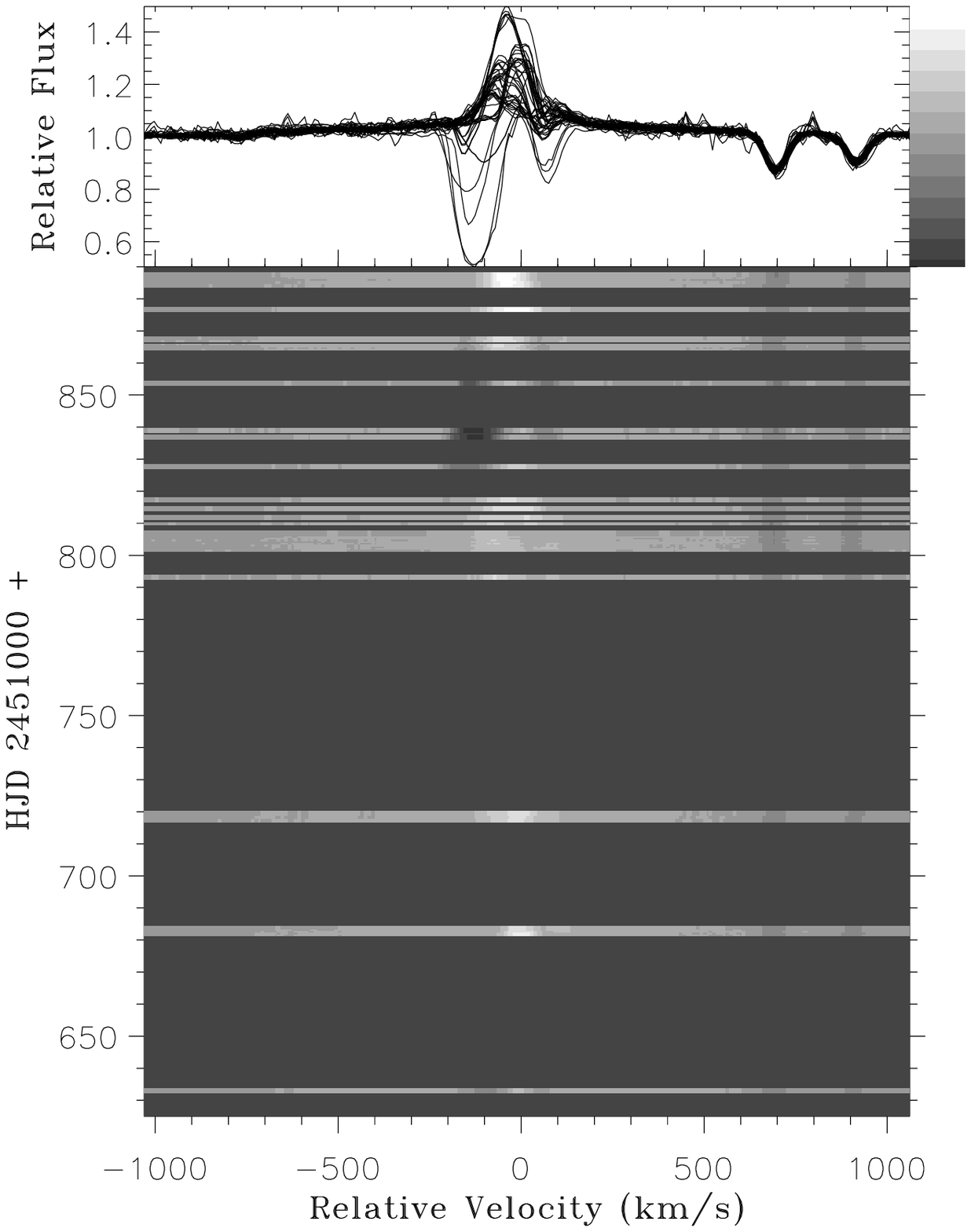}}
\end{minipage}
\hfill
  \caption{|The 1998 (left), 1999 (middle)) and  2000 (right) 
time-series of \Ha shown as one-dimensional plots (top) and 
two-dimensional grey scale images (bottom). All spectra have 
been corrected for \Vsys=-12~\kms. Velocity scale given with 
respect to the rest wavelength of \Ha.  
   }
   \label{ha2d_set}
   \end{figure*}

The estimated photometric period is somewhat larger but 
still consistent (within 3$\sigma$) with the 11.7~day period 
variation in radial velocity of C~{\sc ii} and 
He~{\sc i}~$\lambda$6678 photospheric lines.  This finding 
strongly suggests that the same physical mechanism (based in 
the stellar photosphere) is  most likely responsible for the 
two phenomena observed, which might be identified as the
signatures of pulsations.\\

However, note that the interpretation of the photospheric 
variability of HD~199\,478 reported here and in \citet{PAM} 
is not straightforward in terms of pulsation. On the one hand, 
radial pulsations are not likely since: first, the period is not stable 
between the observing runs carried out in different years 
and second, with only one exception, the estimated 
periods are longer than the radial fundamental pulsational 
period, $P_{\rm {rad,fund}}\sim$8 days, as derived by \citet{MV}.
On the other hand, the irregular character of this variability is 
quite similar to that observed in other late B SGs and A-type 
stars \citep{Kaufer97}. A possible origin for these variations, 
at least for stars with \Mstar  $\le$ 40~\Msun, is the action 
of non-radial oscillation modes excited by the opacity mechanism. 

In this respect, we note that: 
\begin{itemize}
\item[i)] a period of about 20~days, as derived from photometric 
and spectroscopic data of HD~199\,478, is fully consistent 
with the value inferred via the period-luminosity 
relation for B-type variables with excited $g$-mode oscillations  
(Fig. 2 in \citealt{Waelkens98}); 

\item[ii)] on the HR diagram with parameters derived 
with FASTWIND, HD~199\,478 elegantly joints 
the group of B-type SGs studied by \citet{Burki} 
for which $g$-mode instability is suggested to 
explain their variability (Fig. 3 of \citealt{Waelkens98}).
\end{itemize}

Therefore, non-radial $g$-mode oscillations might explain the 
photospheric variability of HD~199\,478. But our data lacks  
evidence for travelling blue-to-red (prograde) features 
within the absorption troughs of the lines, which normally indicate
non-radial pulsational behaviour.

Clearly, very extended time-series datasets are requisite for 
extracting reliable long period signals from the irregular 
absorption line changes which characterise B SGs. 
These targets lend themselves particularly to modest-sized 
robotic telescopes equipped with high-resolution spectrographs.

\section{Wind variability }
\subsection{\Ha monitoring campaigns in 1999 and 
2000}
\label{ha_datasets}

In Figure~\ref{ha2d_set} the \Ha time-series for 1999 and 
2000 are shown as two-dimensional gray-scale 
images. Above each of the velocity-time frames the 
corresponding one-dimensional spectra are plotted to provide 
a visual assessment of the size of the fluctuations at 
each velocity bin. Gaps between observations, if equal or 
larger than 1.0 day, are represented by black bands. All 
spectra have been corrected for the systemic velocity, 
\Vsys~=~-12~\kms. The zero point in velocity 
corresponds to the rest wavelength of \Ha. A similar plot, 
illustrating the \Ha time-series obtained in 1998 
(from \citealt{MV}), is also provided for completeness.

Figure~\ref{ha2d_set} demonstrates that the \Ha 
profile of HD~199\,478 is 
strongly variable, exhibiting a large diversity of profile 
shapes and behaviour patterns. In particular, and as also 
noted by \citet{MV}, in June-July, 1998 as well as during the 
first two months of 1999, the profile appeared fully in 
emission evolving from a double-peak morphology with a 
blue component that is stronger than the red one, to 
a single-peaked feature centered almost at the rest frame. 
Some hints about the subsequent development of this feature to 
the red seem also to be present. Three such cycles have been 
identified by \citet{MV} (see left and middle panels of 
Figure~\ref{ha2d_set}): the first - between HJD~2450\,968-982; 
the second - between HJD~2450\,998-1009 and the third - 
between HJD~2451\,178-189.  Another cycle taking 
place between HJD~2451\,217-247 can be now be easily recognised
thanks to the new observations in March 1999. 
This finding implies that the variability pattern described 
above is relatively stable (over at least 9 months) with a
characteristic time-scale of about 15 days and a possible 
re-appearance after one month or longer.

A new variability pattern is revealed by the
 1999 and 2000 observations, where \Ha appears not only in emission,
 but also in partial or complete absorption.
 In particular, on  HJD~2451\,293 
(April 24, 1999) in addition to the blue-shifted emission 
(\Vr=-75\kms) a slightly red-shifted absorption feature
(\Vr=+40~\kms) has appeared giving rise to a reverse P~Cygni 
profile. The latter persisted for at least 8 days, 
growing slightly stronger in intensity. 

About five months later, namely on HJD~2451\,439 (Sept. 17), 
\Ha is seen fully in emission again, though with a weak 
dip at about +60 \kms, which makes the profile appear 
double-peaked 
with a blue component being much stronger than the red one. 
This configuration was preserved for at least 7 days, 
i.e. up to HJD~2451\,449. On HJD~2451\,475 the 
absorption dip is missing 
but one month later it appears again, stronger than before, 
and persists for at least 6 days (between HJD~2451\,509-515) 
during which time the profile again looks like a reverse P Cygni.
Interestingly, the second appearance of the dip is exactly 
at the same position as the first one suggesting the same
physical origin for both  events.

The 2000 observations importantly
revealed (right panel of Figure~\ref{ha2d_set}) the presence of 
another unusual event during which \Ha changes suddenly and 
drastically from pure emission to pure absorption and back to 
pure emission. By chance, the distribution of the available 
observations in time was  quite good allowing the development 
of this spectacular event to be followed in more detail.
\begin{figure}
\begin{minipage}{7.5cm}
\resizebox{\hsize}{!}
{\includegraphics{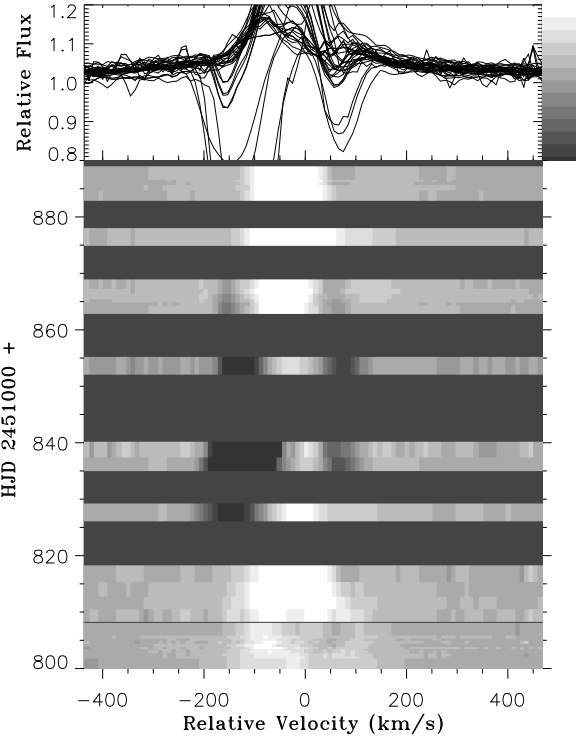}}
\end{minipage}
\hfill
\caption{HVA event observed in the \Ha data of HD~199478 in 2000.  
   }
   \label{HVA_2d}
   \end{figure}

In Figure~\ref{HVA_2d} one can see that before the onset of the 
high-velocity absorption (HVA) event, \Ha appeared fully in emission 
developing from a 
double-peaked to a single-peaked morphology and strengthening slightly 
with time (HJD~2451\,792-816). On HJD~2\,451\,827, in addition to 
the emission a localised high-velocity (\Vr=-150~\kms) absorption 
extending from -68 to -250 \kms is present making the profile 
appear P~Cygni-like. Over the next 9 days the P Cygni feature 
evolves into double absorption with central emission where
the blue component is significantly stronger and wider than 
the red one. Two weeks later, (HJD~2\,451\,863) the morphology 
of the profile is still the same though the blue component is 
weaker and narrower while the red one has apparently 
strengthened becoming somewhat wider. Subsequently 
the two absorptions are fading in parallel and disappear completely 
on HJD~2\,451\,877. 
\begin{figure}
\begin{minipage}{4.3cm}
\resizebox{\hsize}{!}
{\includegraphics{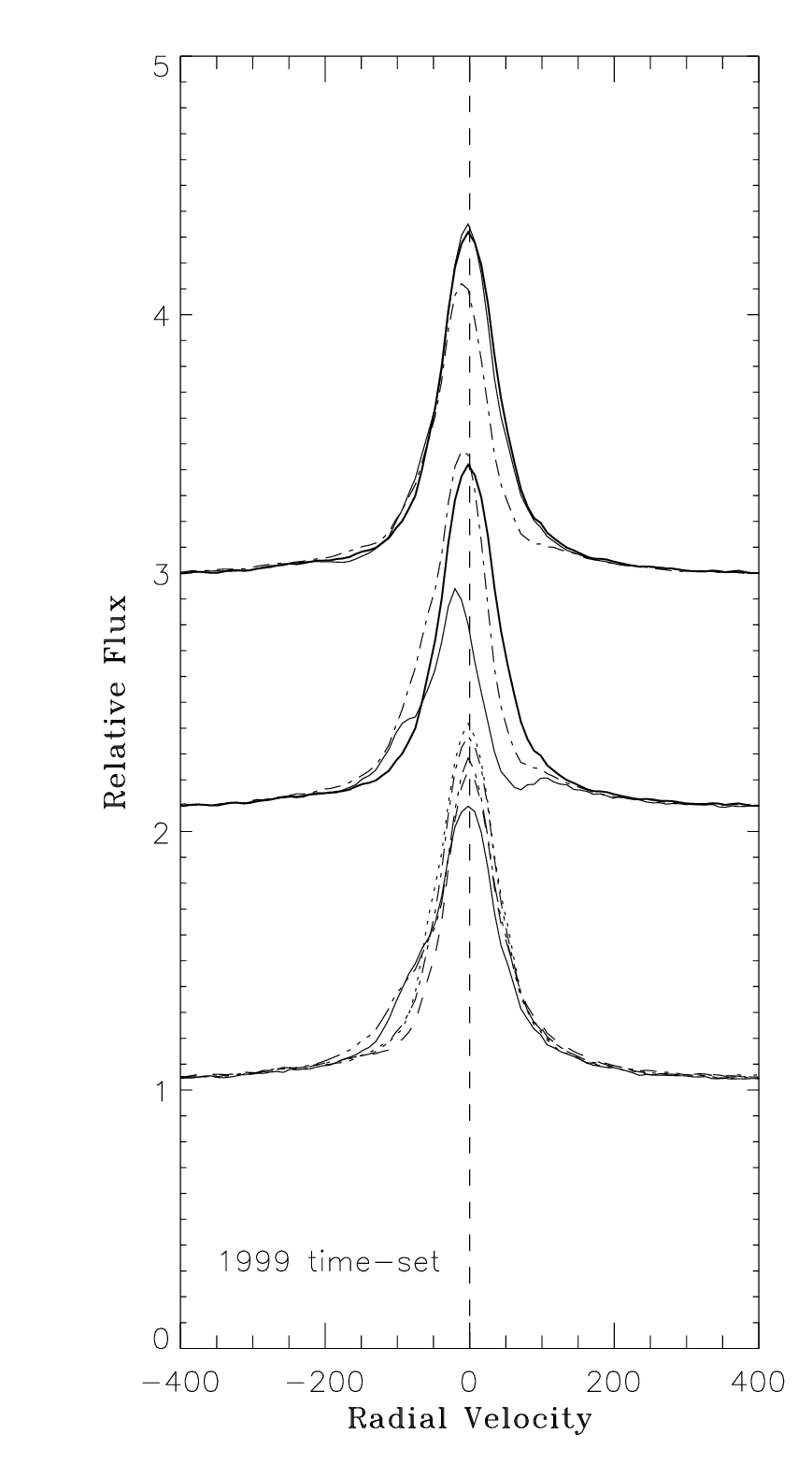}}
\end{minipage}
\begin{minipage}{4.3cm}
\resizebox{\hsize}{!}
{\includegraphics{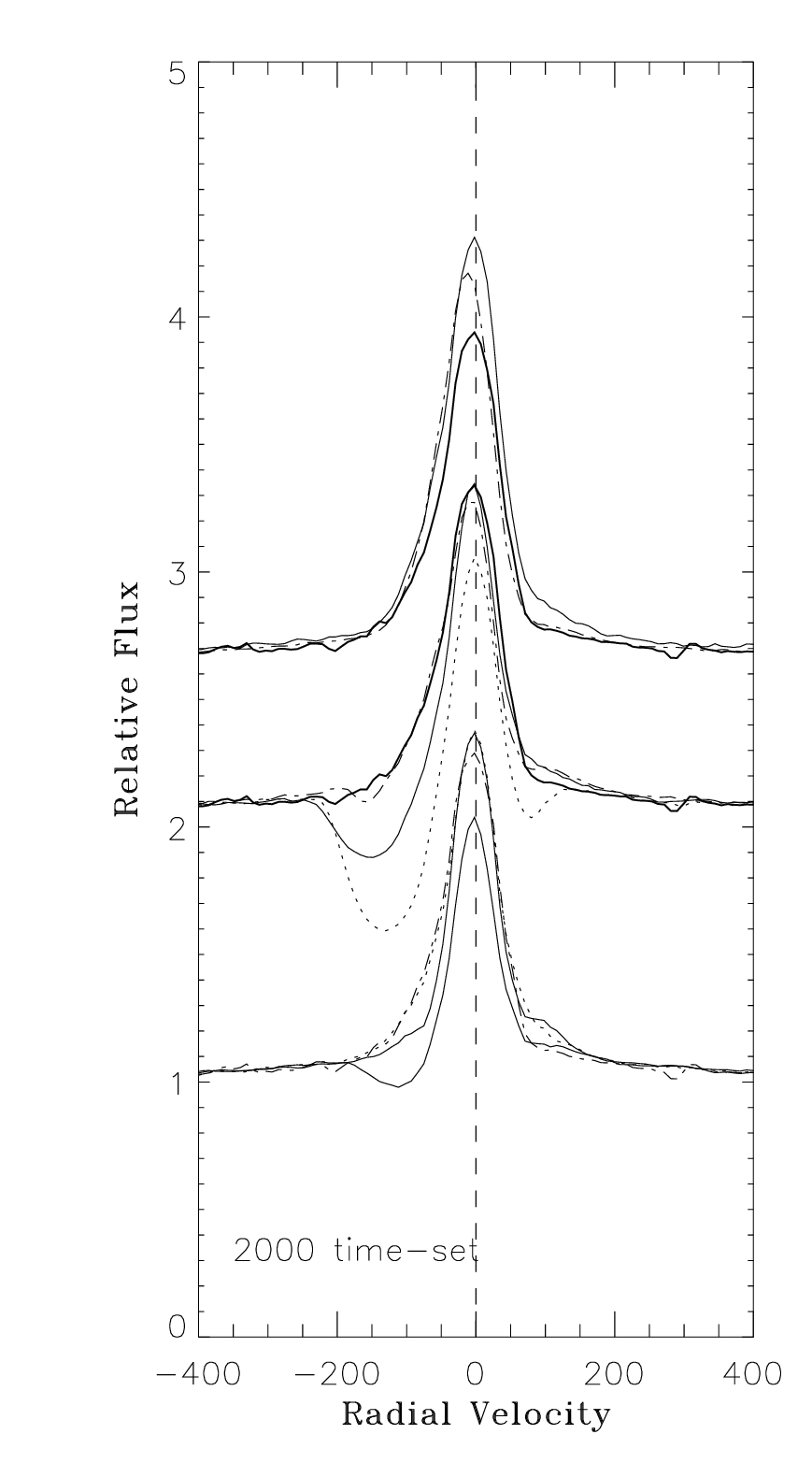}}
\end{minipage}
\caption{Top: \Ha profile of HD~199\,478 illustrating the behaviour 
of pure wind contribution to the line (in velocity space).
   }
   \label{Ha_norm}
   \end{figure}
Our observations further suggest that the absorption 
event seen in the \Ha time-series of HD~199\,478 between HJD~2451\,827 -- 
876 may not be unique. Indeed, readers should
note that on HJD~2451\,632 
(see right panel of Figure~\ref{ha2d_set}) \Ha has also 
appeared as a double absorption feature. Unfortunately, due to
poor temporal coverage the time development of this feature cannot 
be followed, but given the similarity in the morphology of 
this profile and the one taken, e.g., on HJD~2451\,853 we are  
tempted to speculate that about 6 months earlier an absorption  
phenomena similar to the one recorded in September - October 
2000 may have occurred in this star. 

\subsection{Wind contribution}

To probe further the nature and origin of 
peculiar emission and HVAs in the \Ha data of HD~199\,478 we 
normalised the observed \Ha profiles to a constant photospheric 
profile \footnote{Such an approximation is legitimate since the  
observed photospheric variability in HD~199\,478 is indeed very weak 
(see Sec. \ref{he1_c2})} computed by means of the FASTWIND code 
with parameters from \citet{MP} (given also in Table~\ref{para}).

In Figure~\ref{Ha_norm} the \Ha profiles from the 
1999 and 2000 data-sets are shown (in chronological order 
from the bottom upwards).  Profiles from runs separated by large gaps 
are grouped together where periods without significant $lpv$ 
are represented by a single averaged profile 
so as not to confuse the figure. 

The following points are immediately apparent from these plots: 

\noindent
$\bullet$ 
Outside the HVA events the wind contribution can be assigned 
to two components: (i) a strong emission feature which is either 
symmetric with respect to the stellar rest frame or shows weak 
blue-to-red 
asymmetry  with a blue wing being more extended and stronger than 
the red one and (ii) localised emission bumps with variable 
position  which more likely give rise to the established $V/R$ variations.

\noindent
$\bullet$ 
During the HVA episodes (right panel, bottom and middle groups) in 
addition to the extended blue-shifted absorption an emission component 
also exists, i.e. the wind does not only absorb but also emits \Ha  
photons, contrary to the cases described by \citet{kaufer96b}, 
where the HVAs are not accompanied by unshifted emission.

\noindent
$\bullet$ 
The red-shifted absorption seen occasionally in H$\alpha$, is more likely 
of wind origin and suggests the presence of matter infall at the 
base of the wind.  

From the properties outlined above one might conclude that the envelope 
of HD~199\,478 
consists of (i) a spherical component where the physical conditions 
favour only processes which produce emission in \Ha and (ii) localised 
large-scale wind structures with matter infall and outflows, where 
emission and/or absorption can originate.

\subsection{Comparisons of high-velocity 
absorption event in HD~199\,478 to those in other late-B SGs}

Comparison of our Figure~\ref{HVA_2d} with similar results from  
\citet{kaufer96a,kaufer96b} showed that the spectacular absorption 
event seen in \Ha of HD~199\,478 is qualitatively similar to the 
HVAs observed in HD~34\,085 (B8 Ia, $\beta$ Ori), HD~91\,619 (B7 Ia) 
and  HD~96\,919 (B9 Ia), though with one exception: in our data-set 
the blue and the red-shifted absorption components do not merge 
to form an extended blue-to-red absorption, as is the case of the 
objects Kaufer et al. studied, but instead occur parallel to each other 
(though we accept the caveat that a more intensive and extended dataset is ideally required). 

\begin{figure}
\begin{minipage}{7.8cm}
\resizebox{\hsize}{!}
{\includegraphics{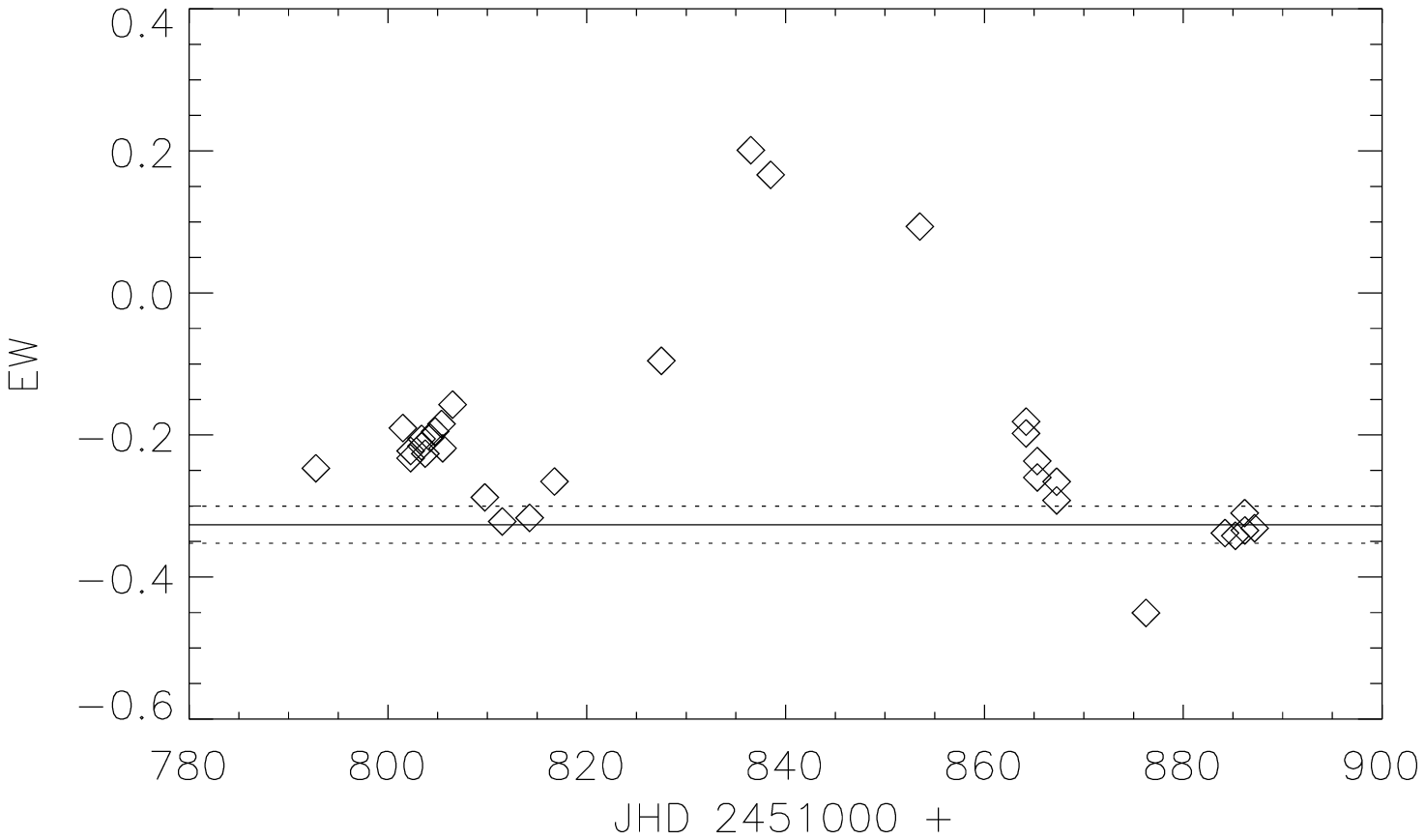}}
\end{minipage}
\\
\hfill
\begin{minipage}{7.8cm}
\resizebox{\hsize}{!}
{\includegraphics{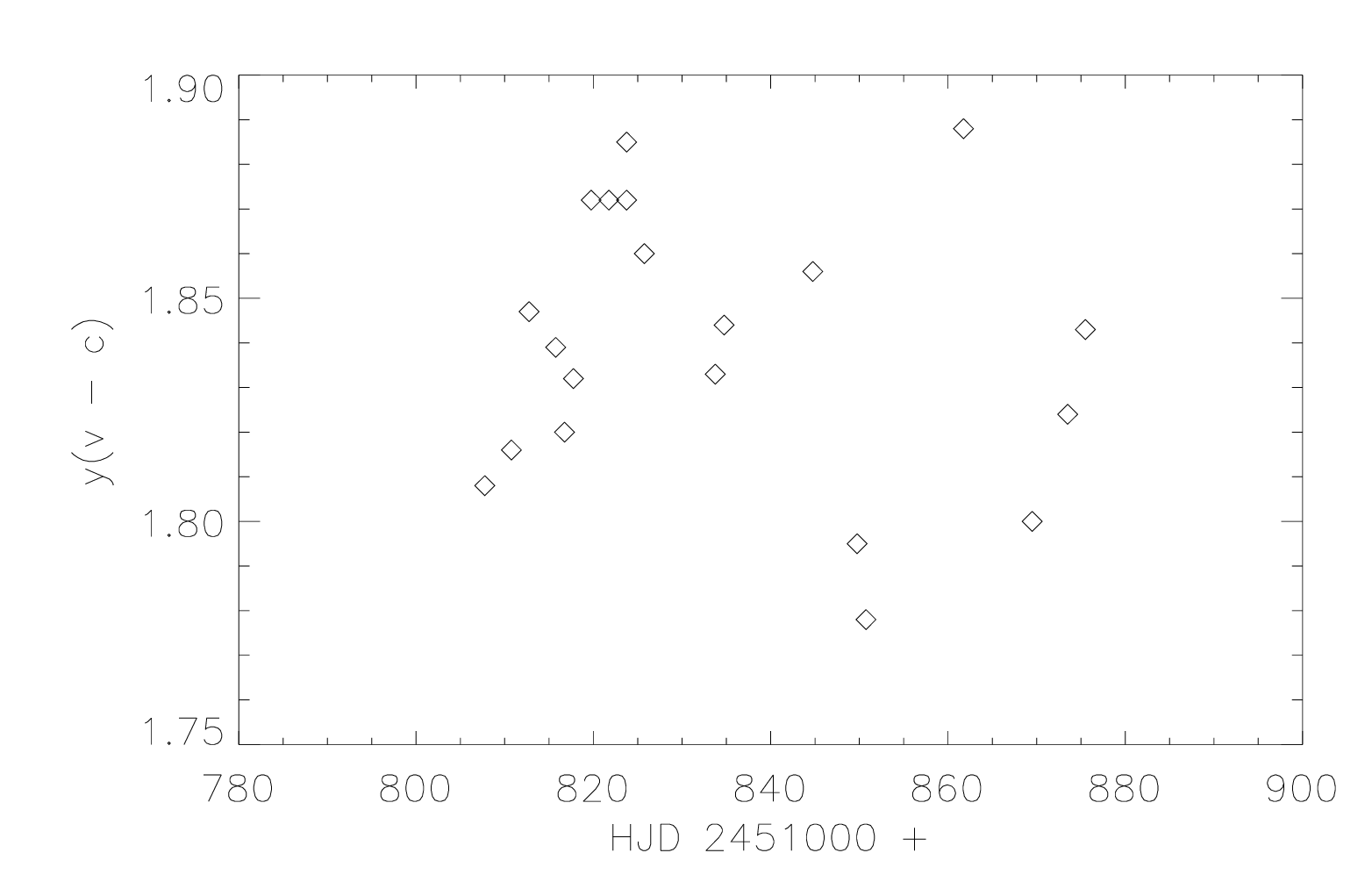}}
\end{minipage}
\caption{Top: Total equivalent width of 
the \Ha line profile of HD~199\,478 
during the 2000 HVA event as a function of time. The horizontal 
lines represent the ''undisturbed'' mean equivalent width of \Ha 
(fully-drawn) and its corresponding standard deviation (dotted).
\newline
Bottom: Time behaviour of the differential y-magnitudes 
during the HVA event 
   }
   \label{HVA_EW}
   \end{figure}

That the spectacular phenomena of HVAs in \Ha have been observed 
so far in 4 late B SGs with peculiar emission in \Ha deserves 
special attention since it might indicate some 
fundamental property of their stellar winds. With this 
in mind we followed \citet{kaufer96b} and 
measured the main properties (such as, e.g., relative intensity, 
position and blue- and red-edge velocities) of the 
\begin{table}
\caption{Properties of HVAs in \Ha observed in 3 late-B 
SGs. Data for HD~34\,085 and  HD~96\,919 are taken 
from \citet{kaufer96b}. All velocities, measured with respect 
to the stellar rest frame, are given in units of the 
corresponding wind terminal velocities (from Table ~\ref{para}).}
\label{HVA_char}
\tabcolsep1.5mm
\begin{tabular}{lrrr}
\hline
\hline
                            & HD~34\,085&HD~96\,919&HD~199\,478\\
\multicolumn{1}{l}{Signature}
&\multicolumn{1}{r}{1994}
&\multicolumn{1}{r}{1995}
&\multicolumn{1}{r}{2000}
\\
\hline
\\
MJD of max. blue depth           &2449493        &2449792         &2451836 \\
Max depth in \% of cont          &   20          & 70             &    49  \\
Vel. of max. depth               & 0.40...0.60   & 0.29...0.43.   &0.37...0.76 \\
Blue-edge velocity               & 0.79...1.20   & 0.43...0.64    &0.68...1.39 \\
Red-edge velocity                & 0.34...0.52   & 0.32...0.48    &{\bf 0.41...0.84}\\
Rise time of $W_\lambda$ [d]     &  11           &  21            & 22 \\
Decay time of $W_\lambda$[d]     &  20           &  46            & 33 \\
Duration of event [d]            &  40           &  90            & 55 \\ 
\hline
\end{tabular}
\end{table}
2000 HVA in \Ha of HD~199\,478 at the time of its maximum intensity. 
Based on the time evolution of the \Ha total equivalent width (measured by 
integrating the flux between 6554 and 6570 \AA) we 
determined the total duration and the rise and decay times of this 
event (Figure~\ref{HVA_EW}, top panel)

The estimates thus derived are listed in Table~\ref{HVA_char}  together 
with similar data for  HD~34\,085  and  HD~96\,919 (from 
\citealt{kaufer96b}). Note that since in our spectra the blue-shifted 
component of the HVA event never merges with the red-shifted one, 
the red-edge velocity (given in Table~\ref{HVA_char} in bold) does not 
refer to the extended blue-to-red absorption (as is the case 
of \citealt{kaufer96b}) but instead corresponds to the red-absorption 
component itself. Note also that due to the large uncertainties in the 
adopted terminal velocities (see next section) the normalised velocities 
sometimes exceed unity.

Compared to similar events in HD~34\,085 and HD~96\,919, the HVA 
seen in HD~199\,478 is of intermediate duration and strength. 
Its time development is roughly consistent with results 
from Kaufer et al.,  which show rise times that are smaller than the 
decay times. \footnote{This result has to be considered with 
caution since the exact time of maximum depth absorption in 
HD~199\,478 is not known with confidence due to limited
time-series coverage.} Thus, 
the duration 
of a HVA event  seems to depend on its maximum strength (stronger 
maximum absorption -- longer duration), while its development in 
time (rising time vs time of decay) appears to be independent of 
this parameter. In addition, the blue-edge velocity and the velocity 
of maximum depth of a HVA event may anti-correlate with its  
strength, i.e. stronger features tend to reach maximum depth 
at lower velocities, being less extended in velocity space than 
weaker ones. Furthermore, and as also noted by \citet{Israel97}, 
the maximum positive velocity of a HVA is always lower than the 
corresponding maximum negative velocity. (Due to the 
limited number of stars these results can only be
regarded as suggestive and they have 
to be confirmed with improved statistics.)

Finally, we note that the photometric behaviour of 
HD~199\,478 during the 2000 HVA event in \Ha provides tentative 
evidence that at the onset of the event the star was about 
one magnitude fainter than at the moment of maximum line 
absorption (Figure~\ref{HVA_EW}, bottom panel).

\section{Stellar and wind parameters of late-B stars which have 
exhibited HVAs in \Ha}
\label{fastwind}

To help understand the nature and origin of the \Ha variability, 
especially the appearance of HVAs, \citet{kaufer96a} have determined 
the fundamental parameters of their sample stars employing: 
\ben
\item[i)] the Azzopardi spectral type--absolute magnitude and
observed \Hgama equivalent width--absolute magnitude calibrations 
\citep{Azzo81}; 
\item[ii)] the Schmidt-Kaler spectral type--bolometric correction 
and spectral type --effective temperature calibrations
\citep{Schmid-K}; 
\item[iii)]  the evolutionary tracks of \citet{Schaller} to derive 
stellar masses;
\item[iv)] the maximum half-width of the TVS of two isolated 
absorption lines to estimate projected rotational velocities;
see also \citep{Reid93}
\item[v)] the UV resonance lines Mg~II$\lambda\lambda$2795,2803 to 
determine wind terminal velocities, \vinf. 
\een

By means of these parameters the authors have subsequently constrained 
the true rotational periods (by means of  $P_{\rm rot}$/sini $i$ 
and $P_{\rm rot,break}$) and evaluated the radial fundamental 
pulsational periods of the stars. The  analysis led 
the authors to suggest that rotation plays an important role 
in determining the properties of the \Ha variability including 
the HVA events.

Prior to the use of currently available state-of-art 
model atmosphere 
codes, the approach used by \citet{kaufer96a} (with its 
well-known weaknesses and uncertainties) was the only one to 
permit the basic parameters of hot stars to be determined. 

The situation has changed drastically since then and  
stellar and wind parameters of hot stars can now be derived
with relatively high precision using the methods of  the 
quantitative spectral analysis. The outcomes of such 
analyses, performed by means of the present day  NLTE, line 
blanketed model atmosphere codes (e.g. CMFGEN \citealt{hil98} 
and FASTWIND \citealt{Puls05}) have unambiguously showed 
that the newly derived stellar and wind parameters  can 
significantly deviate from their earlier determinations 
(e.g., \citealt{martins05} and references therein for O stars 
and \citealt{crowther06,MP, Sam} for B stars)

With this in mind and given the limited number of late B SGs  
with reliably determined stellar and wind parameters (see 
\citealt{MP} and references therein), we decided to re-determine 
the parameters of the Kaufer et al. late-B SGs with HVAs in 
\Ha, using optical spectra kindly provided by Otmar Stahl 
and employing the one of the latest version of  the FASTWIND code. 
This way a homogeneous data base for late B SGs sharing similar 
empirical properties in \Ha would be created, which 
might be easily extended in the future.
\begin{table*}
\caption{Stellar and wind parameters of HD~34\,085 ($\beta$ Ori) 
and HD~96\,919 
as derived in the present study employing the FASTWIND code. 
Numbers in brackets refer to similar data from \citet{kaufer96a}. 
Estimates for HD~199\,478 are taken from \citet{MP}. Effective 
temperatures are given in kK, velocities in \kms, time periods 
in days. Estimates for \Mdot and \logwm are lower limits. }
\label{para}
\begin{tabular}{lllllllllll}
\hline 
\hline
\multicolumn{1}{l}{star}
&\multicolumn{1}{l}{sp}
&\multicolumn{1}{l}{distance}
&\multicolumn{1}{l}{\MV}
&\multicolumn{1}{l}{\Teff}
&\multicolumn{1}{l}{\logg}
&\multicolumn{1}{l}{\Rstar/\Rsun}
&\multicolumn{1}{l}{\logl}
&\multicolumn{1}{l}{\Mstar/\Msun}
&\multicolumn{1}{l}{log \Mdot}
&\multicolumn{1}{l}{\logwm}
\\
\hline
\\
HD~91\,619 &B7Iae &2.51 &-7.00(-7.99) &13.9(12.2) &1.85(1.75) &63(114) &5.13(5.42) &11(27) &-6.92$\pm$0.32 
&27.18$\pm$0.33\\
HD~199\,478 &B8Iae &1.84 &-7.00       &13.0       &1.70       &68      &5.08       &9      &-6.73...-6.18  
&27.33...27.88  \\
HD~34\,085 &B8Iae &0.50  &-8.31(-7.77)&12.5(11.2) &1.70(1.67) &129(116)&5.56(5.28) &31(23) &-6.47$^{\rm +0.27}_{\rm -0.15}$ 
&27.75$^{\rm +0.27}_{\rm -0.20}$\\
           &   &0.24$^{a}$&-6.70   &           &           &61      &4.92       &7      &-6.96$^{\rm +0.37}_{\rm -0.30}$ 
&27.10$^{\rm +0.37}_{\rm -0.32}$\\
HD~96\,919 &B9Iae &&-7.0$^{*}$(-7.97)&11.0(10.3) &1.50(1.50) &71(141) &4.82(5.30) &6(23)  &-7.10$^{\rm +0.25}_{\rm -0.13}$ 
&27.03$^{\rm +0.27}_{\rm -0.21}$ \\
\hline
star         &\Vsys&\vmac&\vmic&\vsini &$V_{\rm break}$&$P_{\rm rot},break$&$P_{\rm rot}/sin i$&\vesc&\vinf  &\\
\hline
HD~91\,619   &-6   &35   & 8   &35(60) &153(212)       &21(27)             &91(96)             &220(261) &170...330 \\
HD~199\,478  &-12  &40   & 8   &41     &134            &25                 &84                 &191      &170...350 \\ 
HD~34\,085   &+18  &35   & 8   &30(55) &184(195)       &35(30)             &218(107)           &262(244) &230...350 \\
             &     &     &     &       &129            &24                 &103                &183      &230...350 \\
HD~96\,919   &-24  &25   & 7   &30(60) &112(176)       &32(40)             &120(119)           &160(220) &250...370 \\
\hline
\end{tabular}
\newline
$^{a}$ - $HIPPARCOS$ distance estimate\\
$^{*}$ - absolute magnitude from the calibration of \citet{HM}\\
\end{table*}  
%

To perform our analysis we followed the strategy outlined in detail by 
\citet{MP}. In particular, effective temperatures, \Teff,  
were estimated from the Silicon ionization balance, fitting 
the Si~II doublet at 4130 \AA\, and the Si~III triplet at 
4552 \AA\, and adopting a solar Silicon abundance (log (Si/H)= -4.45 by 
number\footnote{According to latest results \citep{Asplund}, the 
actual solar value is slightly lower, log (Si/H) = -4.49, but 
such a small difference has no effect on the quality of the 
line-profile fits.}, cf. \citealt{GS98} (and references
therein), and a microturbulent velocity, \vmic, appropriate 
for the corresponding spectral type \citep{MP}. 
Since in all objects the blue wing of \Hgama seems to 
be affected by blue-shifted wind emission (similar to the one 
seen in \Ha) surface gravities, \logg, were derived fitting 
the wings of \Hdelta.  The accuracy of these estimates is 
$\pm$500~K in \Teff and $\pm$0.1 in \logg.  

\paragraph{Projected rotational and macroturbulent velocities, 
\vsini\, and \vmac,} $-$ these parameters were determined from Mg~II 
 $\lambda$ 4481  
employing the Fourier technique developed by \citet{simon}. 
Since this method provides only rough estimates of \vmac the
latter have been  additionally adjusted during the fitting 
procedure (if necessary) to improve the quality of the fits. 
Radial velocities from \citet{kaufer96a} have been
adopted. 

\paragraph{Stellar radii, \Rstar} $-$ These parameters were determined 
from the derived effective temperatures and de-reddened absolute 
magnitudes (see, e.g., \citealt{Kudritzki80}). The latter 
were calculated using standard extinction law with $R = 3.1$ 
combined with:  (i) visual magnitudes,  $V$, and $B-V$ colours 
from  the $HIPPARCOS$ $Main$ $Catalogue$  ($I/239/hip_{main}$); 
(ii) intrinsic colours $(B-V)_{\rm 0}$ = -0.03  from 
\citep{FG}, and (iii) distances collected from various sources 
in the literature.  

In particular, for HD~91\,619, a member of Car~OB1 
association, a distance of 2.51~kpc as provided by  
\citet{Humph} was adopted. For HD~34\,085 ($\beta$~Ori) the 
situation is a bit more complicated.  As a member of Ori~OB1  
this star should be situated at about 0.5~kpc  \citep{Humph}. 
Its possible membership of the $\tau$ Ori~R1 complex 
\citep{HJ82} reduces the  distance to about 0.36~kpc, while  
the $HIPPARCOS$ distance estimate is 0.24~kpc. 
\footnote{Below 0.5~kps the $HIPPARCOS$ distance estimates 
are generally accepted as reliable.} Thus, for $\beta$~Ori 
we provide two entries as upper and lower limits to 
the distance to account for all possibilities. For 
HD~96\,919, which does not belong to any cluster or association, 
an absolute magnitude according to the calibration of \citet{HM} 
was adopted. We quote typical uncertainties  of $\pm$500~K 
in our \Teff estimates and of $\pm$0.4 in \MV (for members 
of associations) to 0.5~mag (for stars with \MV from 
calibration)  \citep{MP}. The error in 
the stellar radius  is dominated by uncertainties in 
\MV and is of the order of $\Delta$log~\Rstar= $\pm$0.08...0.10, 
i.e., less than 26\% in \Rstar.

\paragraph{Luminosities, \logl, and stellar masses, \Mstar,} $-$
these values 
were determined from the corresponding \Teff and \Rstar 
values and  the ``true'' surface gravities, respectively. 
\footnote{``True'' gravity results from the observed gravity 
corrected for the centrifugal acceleration (=(\vsini) 
$^{\rm 2}$)/ \Rstar). Due to the lower \vsini of the 
sample stars this correction is generally small, 
between 0.01 to 0.03.} The typical uncertainties of these
estimates are $\Delta$\logl=$\pm$0.17 to 0.22 and 
$\Delta$log\Mstar=$\pm$0.19 to 0.22.
 
\paragraph{Terminal wind velocities,\vinf,} $-$ values for $\beta$ Ori 
and HD~96\,919 have been determined from the blue-edge of 
the Mg~II resonance lines at $\lambda\lambda$~2795, 2803 
by \citet{kaufer96a}. However, the authors note that due to 
the absence of sharp blue edges, their estimates have to be 
considered as lower limits only. Thus, in these two cases 
we adopted the  Kaufer et al. estimates of \vinf 
(i.e. -230~\kms for HD~34\,085 and -250~\kms for HD~96\,919) 
but assumed an asymmetric error of +50\%   to allow for a 
rather large uncertainty towards higher values. For 
HD~91\,619, since no \vinf estimate was found in the 
literature, we followed \citet{MP} and adopted 
\vinf = \vesc = 220~\kms assuming an asymmetric error 
of -25/+50\%.

\begin{figure*}
\begin{minipage}{4.2cm}
\resizebox{\hsize}{!}
{\includegraphics{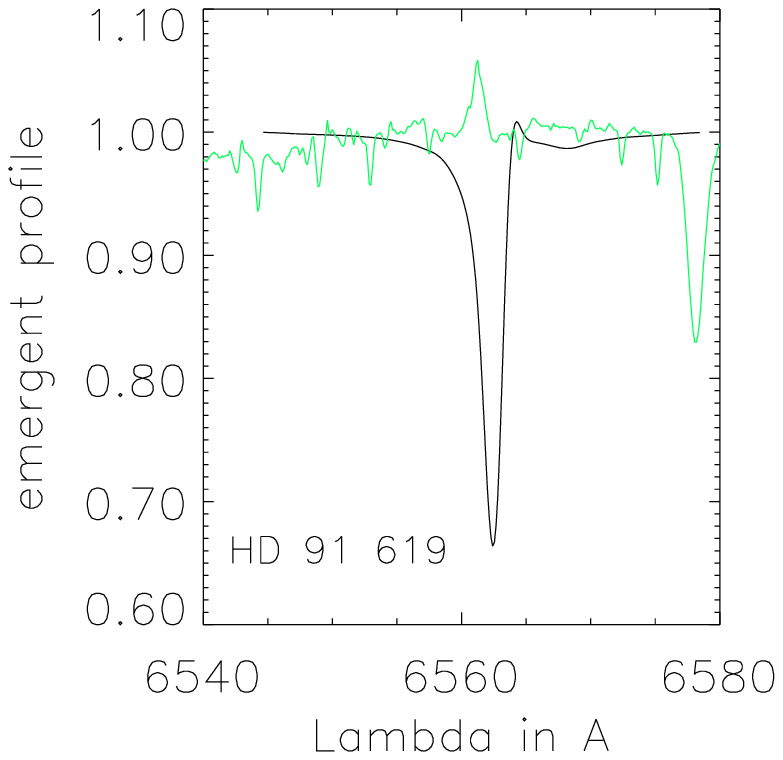}}
\end{minipage}
\begin{minipage}{4.2cm}
\resizebox{\hsize}{!}
{\includegraphics{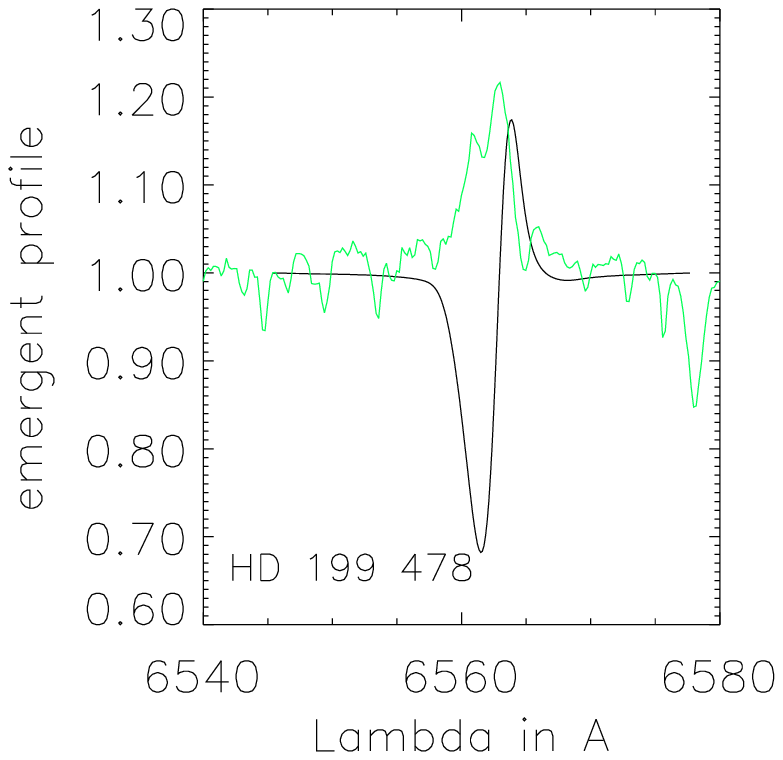}}
\end{minipage}
\begin{minipage}{4.2cm}
\resizebox{\hsize}{!}
{\includegraphics{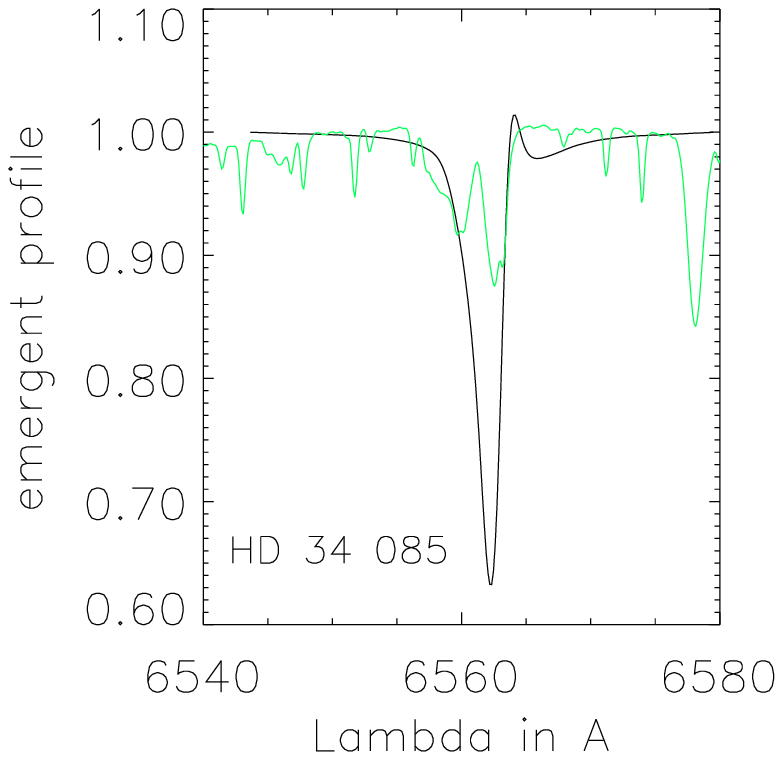}}
\end{minipage}
\begin{minipage}{4.2cm}
\resizebox{\hsize}{!}
{\includegraphics{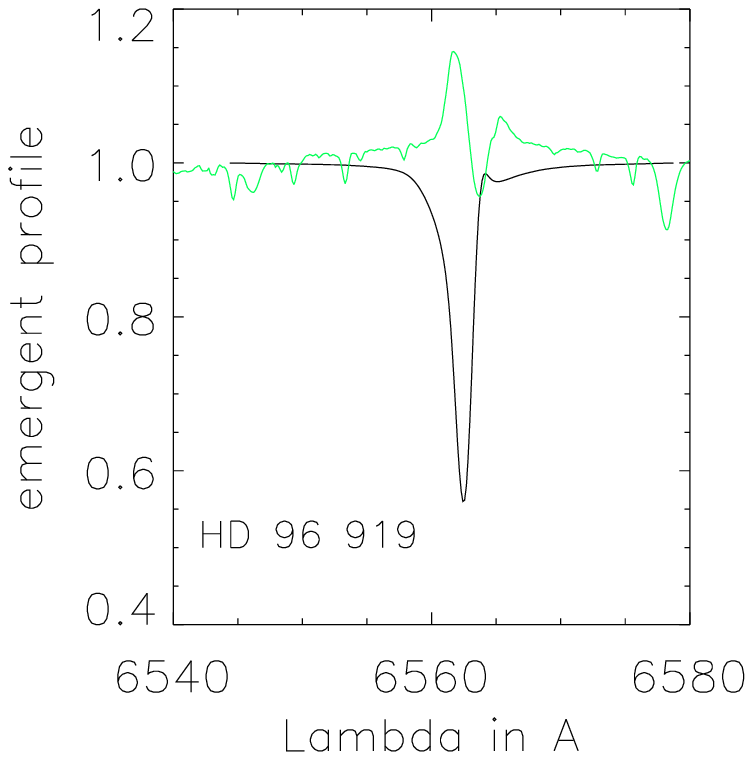}}
\end{minipage}
  \caption{Examples of typical \Ha profiles observed in the four 
sample stars which have been recognised to show HVA events. 
Over-plotted are synthetic profiles accounting for the contribution 
of photospheric absorption plus spherically symmetric smooth 
wind emission (for more info see text.)
}
   \label{model}
   \end{figure*}

\paragraph{Mass-loss rates, \Mdot, and  velocity 
exponent $\beta$} $-$ In the case of strong $undisturbed$ winds 
(\Ha in emission) \Mdot and $\beta$ can be estimated, with 
relatively high precision, from the best fit to the red 
wing and the peak emission of  the \Ha profile, respectively. 
Indeed, \Ha is in emission in the spectra of HD~91\,619 and HD~96\,919
available to us. However, and as also shown 
in Figure~\ref{model}, such profiles cannot be reproduced in 
terms of spherically symmetric smooth wind models since at 
this temperature regime the models predict profiles in 
absorption partly filled in by wind emission. Due to this reason 
only lower limits to \Mdot are derived for our sample stars, 
with $\beta$ ranging from  0.8 to 1.5, except 
for HD~96\,919 where an upper limit to $\beta$ of 1.3 was
adopted (see \citealt{MP}). 
Under these circumstances any excess emission seen in \Ha should 
be attributed either to deviations from spherically symmetric 
wind approximations and/or to processes different from 
recombination (see sect.~\ref{disk}).  

The errors in our \Mdot estimates (actually in log$Q$) accumulated
from uncertainties in $\beta$ and in \Rstar are typically 
less than 0.30~dex. Having \Rstar, \Mdot and \vinf thus 
determined we finally calculated the modified wind momentum,  
$D_{\rm mom}$=$Q$\vinf\Rstar$^{\rm 2}$, with a typical 
error $\Delta$\logwm between 0.13...0.35~dex.

\subsection{Comparisons to parameters derived in other investigations.}

Stellar properties derived in our analysis, together with 
the data for HD~199\,478 (from \citealt{MP}), are listed in 
Table~\ref{para}. Estimates from \citet{kaufer96a} are also provided 
(numbers in brackets) for comparison. 

Compared to similar data from Kaufer et al.:

\noindent
$\bullet$ 
Our \vsini estimates are about a factor of 0.5 lower. This result is 
easy to interpret since our estimates account for 
the effects of  macroturbulence. Contrary to what might be 
expected, the lower values of \vsini did not result in significantly 
different values of the rotational period (see Table~\ref{para}). With our approach the decrease in \vsini is almost 
completely compensated by an increase in \Rstar (see below). 

\noindent
$\bullet$ 
Our \Teff and \logg estimates are surprisingly similar to those in 
Kaufer et al.  This finding 
indicates that in these particular cases the effects of line blanketing
are not too large (as also shown by \citealt{MP}) and that in the 
Kaufer et al. approach the effects of the larger stellar masses 
have been to a  large extent compensated by the larger radii. 

\noindent
$\bullet$ 
Our \Rstar and \logl are significantly lower than those of Kaufer et al. 
The reason for this is the large difference in the adopted absolute 
magnitudes: according to our estimates the three stars are about 
one magnitude fainter than estimated via the Azzopardi calibrations 
(apart from  HD~34\,085 ($\beta$ Ori),  first entry, where there
is an alternative possibility 
of \MV  being half a magnitude brighter). 
Such large differences in \MV cannot be solely explained by 
differences in the methods used, but might indicate instead problems with 
the Azzopardi calibration (see also \citealt{MP}). 

\noindent
$\bullet$ 
In general, our estimates of \Mstar tend to be lower (by about 0.39 to 
0.58~dex) than those estimated by Kaufer et al. via the evolutionary 
tracks (apart from the first entry for $\beta$ Ori, where a value 
of 0.12~dex  larger was derived). This finding is consistent with similar 
results 
from previous investigations \citep{crowther06,Trundle05,MP} which also 
indicate a mild  ``mass discrepancy'' for B SGs.
Indeed, the differences 
established here are somewhat larger, but in their study 
Kaufer et al. used evolutionary tracks which do not take into 
account the effects of stellar rotation, e.g., \citet{Schaller}, while in 
all other studies evolutionary tracks from \citet{MM00} (with stellar 
rotation) have been used instead.

On the other hand, the stellar radii and luminosities derived in the 
present study  are consistent with similar results from  \citet{MP}.
In particular, they fit very well (within the existing scatter) 
the log~Teff--\Rstar and log~Teff--\logl distributions derived by 
these authors for Galactic B SGs. The only real outlier is 
HD~34\,085 ($\beta$~Ori), first entry, where larger deviations ($\sim$ 0.6~dex 
in \logl and a factor of 2 in \Rstar) have been established.
Since the \logg-value of this star is typical for a late SGs 
we suggest that
the photometric distance provided by \citet{Humph} is  overestimated.

Finally, strong agreement (within the corresponding errors) 
was found between our \Teff, \logg  and \vmic estimates 
of HD~34\,085 ($\beta$ Ori) and 
those  derived by \citet{Przb06} via a hybrid non-LTE 
technique (\Teff=12\,000$\pm$200~KK, \logg=1.7$\pm$0.1, \vmic=7$\pm$1~\kms)
Good agreement was also found with the results from \citet{Israel97} 
for HD~34\,085 (\Teff=13000, \logg=1.6 and \vmic=7) obtained via  
the NLTE unblanketed plane-parallel hydrostatic code TLUSTY 
\citep{Hubeny}. The latter indicate that at these temperatures the 
effects of line blocking/blanketing are small as also found by 
\citet{MP} and that in the particular case of this star the wind
effects also seem to be minimal \footnote{(Although the somewhat lower 
value of \logg of Israelian et al. more likely results from the neglect 
of the wind effects)}.

\section{Discussion and conclusions}
\label{disk} 

Extensive monitoring campaigns of several late-B SGs, 
namely HD~199\,478 (present study as well as \citealt{MV})
and  HD~91\,619, HD~34\,085 and HD~96\,919 \citep{kaufer96a,
kaufer96b, Kaufer97, Israel97}, indicate that their \Ha 
profiles exhibit quite similar peculiarities, consisting of 
a double-peaked emission with V/R variation and occasional 
episodes of strong absorption indicating simultaneous mass 
infall and outflows. Such line signatures cannot be reproduced 
by conventional (i.e. non-rotating, spherically 
symmetric, smooth) wind models, which instead 
predict profiles in absorption partly filled in by emission 
for SGs at this temperature regime. 

Discrepancies between observed and predicted \Ha profiles 
have been also established for many O and $early$ B SGs, 
where this finding  was  usually interpreted as an indication 
for deviations from the adopted spherically symmetric, 
smooth wind approximations (e.g., \citealt{Morel, Markova04, 
Markova05} and references therein).

Following this reasoning, axially symmetric envelopes,
modulated, at least in the inner parts, by co-rotating weak 
magnetic structures have been assumed to explain the 
appearance and kinematical properties (e.g. double-peaked 
morphology with $V/R$ variations) of the peculiar \Ha 
emission in the spectra of the four late-B SGs noted above 
\citep{kaufer96a, MV}.   

In addition, to account for the sudden appearance 
of HVAs in \Ha and their development in time (e.g. the 
fast rise over a large velocity range, the lack of 
unshifted line emission, and the mooted re-appearance 
over a rotational time-scale), episodic and azimuthally extended, 
density enhancements in the form of co-rotating spirals 
rooted in the photosphere \citep{kaufer96b} or closed 
magnetic loops similar to those in our Sun \citep{Israel97} 
were also suggested. 

It is generally mooted that non-radial 
pulsations (NRPs) and surface magnetic spots may equally 
be responsible for creating large-scale inhomogeneities 
in hot star winds \citep{Full}. However, despite some progress 
(e.g., \citealt{Kaufer06}) no convincing evidence of 
a direct relation between the time-scale of a given 
cyclical (wind) $lpv$ and the predicted time-scale of recurrent 
surface features due to a specific pulsation mode, has been 
derived to date (see \citealt{townsend07} and references 
therein). 

Note in the 
particular case of the four stars discussed here, 
non-radial pulsations due to $g$-modes oscillations were 
suggested to explain absorption $lpv$ in their 
spectra \citep{Kaufer97, MP}. This possibility is partially 
supported by the present results, which indicate that on the 
HR diagram, and for parameters derived with FASTWIND, 
these stars fall exactly in the region occupied by known 
variable B SGs, for which $g$-modes instability was 
suggested. Also, the photometric variability of 
HD~199\,478 seems to be consistent with a possible origin 
in terms of $g$-mode oscillations \citep{PAM}. Thus, it seems 
very likely that the four late-B SGs in our sample are 
non-radial pulsators. Although no clear evidence of any 
causality between photospheric and wind (as traced by H$\alpha$)  
variability has been seen so far for these objects 
(present study as well as \citealt{Kaufer97}), one might 
speculate that their winds are perturbed due to pulsational 
instability, with specific signatures seen in the 
behaviour of H$\alpha$.

An alternate possibility is that magnetic fields could be 
responsible for the appearance of 
large-scale  structures and wind asymmetries in hot stars. 
In particular, magneto-hydrodynamical (MHD) simulations 
for stars with moderately strong rotation, and for stellar 
and wind parameters typical for O and early B SGs (plus 
a magnetic dipole aligned to the stellar rotation) showed that 
depending on the magnetic spin-up, an equatorial compression, 
dominated by radial $infall$ and/or $outflows$ can be created,
with no apparent tendency to form a steady, Keplerian disk
\citep{OD03, DOT08}.

Indeed, due to the lack of strong convection zones associated 
with hydrogen recombination, normal (i.e. without any chemical 
peculiarities) hot stars are generally not thought to be 
magnetically active. However, theoretical considerations 
(e.g. \citealt{CM}) supported by more recent observations 
\citep{Bychkov, Hurbig07} indicate that this may 
not necessarily be true and that relatively strong, stable, 
large-scale dipole magnetic fields are present in different 
groups of B stars (e.g. SPB, Be,  $\beta$ Cep  itself etc.) 

Thus, it seems likely that in at least some hot stars magnetic 
fields can be an alternative source of wind perturbations and 
asymmetries. The potential role of magnetic fields in B SGs 
remains intriguing, especially because it might provide a clue to 
understand the puzzling problem of the simultaneous presence of 
red- and blue-shifted absorption in \Ha profiles of the four 
late-B SGs discussed here.
\begin{table}
\caption{Magnetic field strength, $B$ (in G), required to get 
an equatorial confinement with simultaneous mass infall and outflows
around each of our targets. The Keplerian, $R_{\rm K}$, the Alfven, 
$R_{\rm A}$, and the escape, $R_{\rm E}$, radii (in units of \Rstar 
above the photosphere) 
are calculated following \citet{OD03}. The HD~34\,085 estimates 
correspond to the 2nd entry in Table~\ref{para}.}
\label{magn_field}
\tabcolsep1.5mm
\begin{tabular}{lrrrr}
\hline
\hline
\multicolumn{1}{l}{Parameters}
&\multicolumn{1}{r}{91\,619}
&\multicolumn{1}{r}{199\,478}
&\multicolumn{1}{r}{34\,085}
&\multicolumn{1}{r}{96\,919}
\\
\hline
\\
$R_{\rm K}$                          &0.37        &0.45          &0.38       &0.42 \\
$R_{\rm A}$                          &0.83..3.73  &0.77..4.59    &0.83..3.79 &1.0..4.13 \\
$R_{\rm E}$                          &3.77        &4.58          &3.81       &4.18 \\
$B$ ($R_{\rm K}<R_{\rm A}<R_{\rm E}$)&5..100      &5..180        &5..105     &5..85\\
\hline
\end{tabular}
\end{table}

Guided by these perspectives, we employed the scaling relations 
given in \citet{OD03} and calculated the Alfven, $R_{\rm A}$, the 
Keplerian, $R_{\rm K}$, and the ``escape", $R_{\rm E}$, radii 
of our targets, using data from Table~\ref{para} and fixing 
the magnetic field strength at the values required to create 
an equatorial confinement. Interestingly, the results 
listed in Table~\ref{magn_field}, show that in all four cases a 
very weak dipole magnetic field can effectively channel the 
wind outflows, leading to the formation of an equatorial 
compression with simultaneous radial mass infall and outflow. 

With this in mind, new MHD simulations for the case of mid/late-B SGs 
have been recently initiated. The preliminary results (private 
communication, Asif ud-Doula) indicate that a pure dipole magnetic 
field of only a few tens of Gauss is indeed required to obtain a 
$cool$ equatorial compression (with mass infall and outflow) 
around a rotating star with stellar and wind parameters as derived 
with FASTWIND for HD~199\,478.  (More detailed information about 
the outcomes of this study will be provided in a forthcoming paper.)

An obvious advantage of the model described above is that it has 
the potential to at least qualitatively account for 
some of the puzzling properties of the \Ha line of our targets. In particular, 
the sudden appearance of red and blue-shifted absorptions might be 
explained if one assumes, that due to some reason the plasma in 
the infalling or outflowing zones of the compression or in both 
of them (during the HVA episodes) can become optically thick  in
the $Lyman$ continuum and L$_{\alpha}$, thus forcing \Ha to 
behave as a resonance line, i.e. to absorb and emit line photons. 
The kinematic properties of the resulting absorption features 
\footnote{(Depending on the size of the \Ha forming region 
emission may not appear in the spectrum.)} are difficult to 
predict from simple qualitative considerations but it is
clear that these properties cannot be dominated by 
stellar rotation, but instead will be controlled by the 
physical conditions inside the compression.

Concerning the interpretation of the peculiar \Ha emission, 
the situation is more complicated since such emission can 
originate from different parts of the envelope under quite 
different physical conditions. For example, one can expect 
that the cool, less dense plasma outside the compression will 
only emit line photons (via recombinations), producing pure 
emission feature(s) in H$\alpha$ .

Also, the cool equatorial compression might contribute to 
the \Ha emission, providing the plasma inside the compression 
can occasionally become optically thin in this line. 
However, note that even a plasma that is optically thick in
L$\alpha$ and Lyman continuum can, under specific 
conditions, produce $pure$ emission profiles in \Ha (e.g. $if$ 
collisions dominate the \Ha formation or $if$ due to some  
reasons the 2nd and 3rd levels of Hydrogen go into LTE 
(\citealt{PP96})). 

Therefore, very weak dipole 
magnetic fields might be responsible for creating wind 
structures in the envelopes of late-B SGs. The models 
derived via MHD simulations seem to have the potential 
to account, at least qualitatively, for some of the 
peculiar characteristics of \Ha in the spectra of our 
targets. However, more detailed quantitative analysis
is required to investigate this possibility further.

 New high signal-to-noise observations to prove/disprove
the presence of weak magnetic fields can help to clarify the 
picture. Of course, due to the low strength of the 
magnetic fields required one cannot expect to detect 
these fields directly but indirect evidence such as e.g., 
the detection of X-ray emission, abundance anomalies, 
specific periodic variations in UV resonance lines, 
interferometric observations (for more information see 
\citealt{henrichs01}), might also be considered. Note that 
a weak longitudinal 
magnetic field of about 130$\pm$20~G\footnote{Given that 
only one spectral line has been used for these measurements 
an error of 20 Gauss seems somewhat unrealistic and likely 
represents  only a lower limit.},  might have been 
detected in HD~34\,085 ($\beta$~Ori) \citep{severny}, but
confirmation is lacking. 

Finally, at the cooler edge of the 
B-star temperature regime pure emission profiles in \Ha 
can be accounted for if one assumes the winds are clumped.
Indeed, a spherically symmetric, clumped wind will mimic 
wind densities higher than the actual ones, thus giving 
rise to strong line emission, similar to that in O stars. 
Such winds may also give rise to wind absorption, providing 
some of the clumps are optically thick in H$\alpha$.  
Detailed numerical simulations and line formation calculations 
are required to discriminate  between the different possibilities.

\acknowledgements{
This work was partly supported by the National Scientific 
Foundation to the Bulgarian Ministry of Education and Science 
(F-1407/2004). NM and RKP are also grateful to Bulgarian Academy 
of Sciences and the Royal Society (UK) for a collaborative 
research grant. SA's contribution was supported by NSF Grants 
AST-0071260 and 0507381 to The Citadel.
}

{natbib}
\end{document}